\documentclass{llncs}
\usepackage{amssymb}
\usepackage{amssymb}
\usepackage{graphicx}
\usepackage[misc]{ifsym}
\usepackage{amsmath}
\usepackage{lineno,hyperref}
\usepackage{subfig}
\usepackage{stmaryrd}
\usepackage{url}
\usepackage{color}
\urldef{\mailsa}\path|{yifan_liu|
\urldef{\mailsb}\path|lipengyu|
\urldef{\mailsc}\path|*zcqin|
\urldef{\mailsd}\path|*taowan}@buaa.edu.cn|
\newcommand{\keywords}[1]{\par\addvspace\baselineskip
\noindent\keywordname\enspace\ignorespaces#1}

\begin{document}

\mainmatter  % start of an individual contribution
{\color{red} }
%\title{Emotion Based Model: A New Approach to Predict the Stock Volatility in Chinese Market}
%\title{Rating Posts with Sentiment Dictionary for Stock Prediction}
\title{Stock Volatility Prediction Using Recurrent Neural Networks with Sentiment Analysis}
%\title{Recurrent Neural Network with External Input of Emotional Indicators for Stock Prediction}

\titlerunning{Stock Volatility Prediction Using RNNs with Sentiment Analysis}

\author{Yifan Liu$^{1}$  \and Zengchang Qin$^{*1}$ \and Pengyu Li$^{1,2}$  \and Tao Wan$^{*3}$  }
\authorrunning{Liu, Qin, Li and Wan }
\institute{$^1$Intelligent Computing and Machine Learning Lab, School of ASEE\\
Beihang University, Beijing 100191, China\\
$^2$School of Mechanical Engineering and Automation\\
Beihang University, Beijing, 100191, China\\
$^3$School of Biological Science and Medical Engineering\\
Beihang University, Beijing, 100191, China\\
%\mailsa,
%\mailsb,
%\mailsc,
%\mailsd\\
$^{*}$\tt{\{taowan,zcqin\}@buaa.edu.cn}
}

\toctitle{Lecture Notes in Computer Science}
\tocauthor{Authors' Instructions}
\maketitle
\begin{abstract}
In this paper, we propose a model to analyze sentiment of online stock forum and use the information to predict the stock volatility in the Chinese market. We have labeled the sentiment of the online financial posts and make the dataset public available for research.
By generating a sentimental dictionary based on financial terms, we develop a model to compute the sentimental score of each online post related to a particular stock.
Such sentimental information is represented by two sentiment indicators, which are fused to market data for stock volatility
prediction by using the Recurrent Neural Networks (RNNs).
Empirical study shows that, comparing to using RNN only, the model performs significantly better with sentimental indicators.
\end{abstract}

\section{Introduction}

In time-series data mining, stock market is notoriously difficult to analyze, even be totally unpredictable based on the famous Efficient Market Hypothesis (EMH). As early as 1900s, Bachelier \cite{Bachelier1900Th} applied statistical methods to analyze stock
data, and found that the mathematical expectation of the stock fluctuation tends to be zero. In the 1970s, Fama \cite{Eugene1970Efficient} formally put forward the EMH, which stated that under the condition of market with complete information, investors couldn't gain more than fifty percent of the profits only with the past price, or simply, no one {could} `beat' the market' {continuously}.
The Random Walk Theory (RWT) proposed by Osborne \cite{Osborne} also suggested same conclusion that the stock prices were unpredictable. But all these theories are based on the same assumption that investors are rational and complete market information is available.

Paul Hawtin\footnote{The founder of Derwent Capital Markets and one of early pioneers in the use of social media sentiment analysis to trade financial derivatives.} once said ``For years, investors have widely accepted that financial markets are driven by fear and greed.'' In the actual market, investors cannot be completely rational.
They may be influenced by their emotions and make impulsive decisions \cite{Rzepczynski}.
Therefore, the basic assumption of EMH and RWT is not impregnable. Many researchers have tried to study the correlation between sentiment and stock market volatility to challenge the classical theories. For example, researchers found that some factors, e.g. weather and sports games, can affect public emotion and also the stock market. The sunny weather and the rising stock index had certain correlations \cite{Hirshleifer}. There would be a significant market decline after the soccer lost \cite{Edmans}. In recent years, the rapid development of social networks (Facebook, Twitter, Weibo) opens a new door to measure the public emotion. Bollen et al. \cite{Bollen} analyzed the text content of daily Twitter by using two mood tracking tools: OpinionFinder (OF) and Google-Profile of Mood States (GPOMS).
%that measures positive vs. negative mood.
%Google-Profile of Mood States (GPOMS) measures mood in terms of 6 dimensions (\emph{Calm, Alert, Sure, Vital, Kind,} and \emph{Happy}) and
The authors then used self-organizing fuzzy neural network (SOFNN) to predict the volatility of the Dow Jones Industrial Average (DJIA).
%tested the stock market using a model called SOFNN to predict the fluctuation of the Dow Jones Industrial Average (DJIA).
By considering the sentimental information from Twitter, the prediction accuracy has raised up by 13\%.
%When only using the past value of DJIA as input, the model has an accuracy of 73.3\% in predicting the daily up and down changes. But when they used the public sentiment from the twitter as an additional input, the accuracy raised up to 87.6\%.
%The results were very encouraging and a few similar research are followed in this direction.
The results were very encouraging and this direction was followed by some other similar research.
Zhang et al. \cite{Zhang} found that a burst of public emotion no matter positive or negative, heralded the falling of the index. There were also some research on the individual stock, Si et al. \cite{Si} proposed a technique to leverage topic based sentiment from Twitter to help predict the stock price while O'Connor \cite{O'Connor} found that the popularity of a brand was much related to related tweets and its stock price.

But there are still some problems remained. First, Twitter users are predominantly English speakers, or even worse the investors of
a particular market may not use Twitter to discuss their finance \cite{Bollen}. Second, popular sentiment analysis dictionary can not entirely measure the emotion of the stock investors \cite{Loughran}. Sprenger et al. \cite{Sprenger} selected tweets which mentioned the company in the Standard \& Poor's 100 index, and labeled the tweets with \emph{buy}, \emph{hold} or \emph{sell} signals. With the labeled training data, they used a Naive Bayes classifier to extract the signals from the tweets automatically and calculated the bullishness through these signals. Finally, they found that a strategy based on bullishness signals could earn substantial abnormal returns.

Above all, a great amount of focus has been placed on the correlation between the investors' sentiment and the U.S. stock market. While limited by the Chinese expression complexity, little attention has been paid to the the relevant research on Chinese stock market. According to the World Federation of Exchanges database\footnote{\url{http://www.indexmundi.com/facts/indicators/CM.MKT.LCAP.CD/rankings}}, Chinese market capitalization ranked the second in 2015 in the world. That's the reason we focus on our study of Chinese stock market. In this paper, based on Sprenger's \cite{Sprenger} approach, we propose a model to study Chinese stock market. The sentiment of Chinese investors are from the East Money Forum\footnote{\url{http://guba.eastmoney.com/}}, which is one of the biggest and specified stock forum in China, but it is not public forums like Facebook or Twitter. Each stock has its individual sub-forum which ensures that most posts from the sub-forums are published by the investors who hold or sell this particular stock. In order to avoid the problem that the ordinary dictionary often makes misunderstanding in recognizing investors sentiment, we use a machine learning method to generate our own dictionary and then calculate the sentiment score of the posts based on the dictionary automatically. To study the correlation between Chinese stock market and Chinese investor sentiment, we propose sentiment indicators for the stock volatility prediction model using the Recurrent Neural Networks (RNNs) to obtain a better performance.

\section{Sentiment Analysis}
Bollen et al. \cite{Bollen} proposed a dictionary-based method for sentiment analysis of the financial contexts. However,  Loughran and Mcdonald \cite{Loughran} found that three-fourths of the words identified as negative by the Harvard Dictionary are not typically considered as negative in financial contexts.
The same problem also occurs in Chinese sentiment analysis. We find that Chinese posts in stock forums have some special expressions containing strong emotions. But these expressions rarely appear in common sentiment analysis dictionary. So in
this research, we first need a practical dataset from which we can obtain
a dictionary of financial words, we also develop a simple but effective tool to generate sentimental weights for the words.

\subsection{Data Processing}
%东方财富的网址在第一节里面给过了%
East Money Forum is one of the most influential Internet financial media in China. It has more than 3000 sub-forums for each individual stock. We randomly select 10 stocks as well as the sub-forum of the posts from $25^{th}$ Sept., $2015$ to $30^{th}$ Sept., $2016$ with a web crawler ``Bazhuayu (means ``Octopus'')''\footnote{\url{http://www.bazhuayu.com}}. Nearly 96000 pieces of posts are obtained, and most of them are short and colloquial. They do not follow any strict syntax but contain strong sentiment. We randomly sampled 3427 stock posts from 10 different stocks to do the manual annotation\footnote{\url{http://dsd.future-lab.cn/members/2016/LiuYFProject/data.xlsx}}. If the post expresses an optimistic attitude towards the stock market and suggests to buy, we label it as positive, otherwise, we label it as negative. We have annotated 2067 negative posts and 1360 positive posts manually.
The original Chinese texts need to be preprocessed by segmentation, and a classical Chinese text segmentation tool called ``Jieba" (Chinese for ``to stutter") in Python\footnote{\url{https://github.com/fxsjy/jieba}} is chosen for this.
 %It is based on a prefix dictionary structure to achieve efficient word graph scanning and build a directed acyclic graph (DAG) for all possible word combinations. Then the dynamic programming is used to find the most probable combination based on the word frequency. Besides, a HMM-based model with the Viterbi algorithm is designed for the unknown words.
%After the segmentation, a post S turns into a collection of words:
%\begin{equation}
%S{\rm{ = \{ }}{{\rm{w}}_1}{\rm{,}}{{\rm{w}}_2}{\rm{,}}{{\rm{w}}_3}...{{\rm{w}}_n}{\rm{\} }}
%\end{equation}
%With the terms and the sentimental weights, we get the dedicated stock sentimental dictionary with mixed-gram. To make a visualization form for the sentimental weight dictionary, we have drawn two word clouds. The word size represents for its sentimental weight. From Fig.\ref{fig:FP} ,We can see that `limit-up' have the highest positive weight while `limit-down' have the highest negative weight.
%\begin{figure}[htb]
%\begin{centering}
%\subfloat[the top fifty negative words]{\begin{centering}
%\includegraphics[width=0.5\textwidth]{negative}
%\par\end{centering}

%}\subfloat[ the top fifty positive words]{\begin{centering}
%\includegraphics[width=0.5\textwidth]{positive}
%\par\end{centering}

%}
%\par\end{centering}

%\caption{\label{fig:FP}Word clouds of sentimental weight dictionary. }
%\end{figure}

\subsection{Polarity Model of Sentiment}
The polarity model of sentiment is trained by a collection of texts labelled only by positive or negative.
Emotional words are extracted and each one has an associated sentiment weight. Weights can be learned from the labeled training dataset \cite{Guo}.
%The model assumes that each term in a post has an associated sentiment weight.
The sum of weighted sentiment scores of all terms determines the sentiment polarity (positive or negative) of the post.
If the sum is greater than 0, it is positive and vice versa.
%We choose a mixed-gram model which is a mixture of both uni-gram and bi-gram to get a better performance \cite{Wang}. So
The sentiment score ${h_{\bf{w}}}({\bf{x}})$  of a given post is computed as follows:
\begin{equation}
\label{equ:score}
{h_{\bf{w}}}({\bf{x}})  = f\left( {\sum\limits_{i = 1}^N {{w ^{(i)}}{x^{(i)}}} } \right) = f({{\bf{w}}^T}{\bf{x}});
1 \le i \le N
\end{equation}
where $N$ is the number of all the terms in the corpus,
a term could be uni-gram or bi-gram model.
% mixed-gram model which is a mixture of both uni-gram and bi-gram to get a better performance \cite{Wang}
${{w ^{(i)}}}$ is the sentimental weight for each term ${{t^{(i)}}}$, ${{x^{(i)}}}$ is the term frequency or tf-idf value of the given term ${{t^{(i)}}}$ .
Function $f( \cdot )$ is a sigmoid function to compress the linear combination of sentimental weight into 0 and 1, and make it smooth.
\begin{equation}
f(z) = \frac{1}{{1 + {e^{ - z}}}}
\end{equation}
By using the logistic regression, the target label $y$ is 1 for positive posts and 0 for negative ones. So that ${h_{\bf{w}}}({\bf{x}})$ represents the probability of a post being positive. If we take the threshold value as $0.5$, the prediction of the sentiment is:
\begin{equation}
{y} = \left\{ {\begin{array}{*{20}{c}}
1&{h > 0.5}\\
0&{h \le {0.5}}
\end{array}} \right.
\end{equation}
Given a training corpus with $M$ text posts, ${{{\bf{x}}^{(k)}}}$ denotes the ${k^{th}}$ $({1} \le {k} \le M)$ post feature value vector. We can derive the cost function and its logarithmic likelihood function based on the maximum likelihood estimation. The loss function $J( \cdot )$ is:
\begin{equation}
J({h_{\bf{w}}}({\bf{x}}),y) = \left\{ {\begin{array}{*{20}{l}}
{ - \log ({h_{\bf{w}}}({\bf{x}}))}&{if}&{y = 1}\\
{ - \log (1 - {h_{\bf{w}}}({\bf{x}}))}&{if}&{y = 0}
\end{array}} \right.
\end{equation}
The average loss for the entire data set is (for $1 \le k \le M$ ):
\begin{equation}
\begin{aligned}
J({\bf{w}})&=  - \frac{1}{M}\sum\limits_{k = 1}^M {J({h_{\textbf{w}}}({\bf{x}}),y)} \\
&=  - \frac{1}{M}\left[ {\sum\limits_{k = 1}^M {{y^{(k)}}\log ({h_{\textbf{w}}}({{\textbf{x}}^{(k)}})) + (1 - {y^{(k)}})\log (1 - {h_{\textbf{w}}}({{\textbf{x}}^{(k)}}))} } \right]
\end{aligned}
\end{equation}
In order to minimize $J({\textbf{w}})$, we can update $\textbf{w}$ using the Gradient Descent algorithm with a learning rate $\alpha$: ${\bf{w}}_{j + 1}^{(k)} = {\bf{w}}_j^{(k)} - \alpha ({h^{(k)}} - {y^{(k)}}){{\bf{x}}^{(k)}}$. The values of the sentimental weight can be obtained. The term with a higher sentimental weight indicates a stronger positive sentiment and vice versa. We use the weights and the corresponding terms to build a sentimental dictionary. The sentiment score of a post can be calculated by weighted sum of weights of all consisting terms based on Eq.(\ref{equ:score}).

%With ten percent of data for test, ninety percent for training, we train the above polarity model and get a training accuracy of 99\% while a test accuracy of 84.4\% for binary classification. However, it has a deficiency that there are more negative posts than positive in the training data. Hence some neutral words may be given a quite low negative weight. If we give a binary label to all the unlabeled posts. The number of the negative posts will be much larger than the positive. To avoid this, we rank the post with a sentimental score instead of a binary label. For an unlabeled post, we match all the terms in this post with the sentiment weight dictionary, and sum the weights to get a sentiment score for this post. If a post is neutral, the sentimental score will be very low. Therefore, it won't influence the sentimental indicators of that day. In addition to this we can keep more sentimental information from the stock posts.
\section{Emotion Model for Stock Prediction }
%In this section, we will discuss how to fuse the sentimental information to
%a a fundamental time series model for stock prediction based on Recurrent Neural Networks (RNN) and design comparison %studies considering the EMM to verify the influence of the information from the forums.

\subsection{Sentimental Indicators}
%%herd mentality从众心理
Some literatures in finance \cite{Sprenger} suggest that individual investors have a herd mentality when they make decisions. For example, if they find that most of people are not optimistic in the outlook of the stock price, they will trade on the advice and move the price. What's more, a larger quantity of the posts on a forum indicates a larger amount of attention which may lead to a severe price volatility. Therefore, we propose an emotion model (EMM) according to the following two important assumptions:
(1) Increased bullishness of stock posts is associated with higher stock price.
(2) Increased posts volume suggests a more substantial volatility.
The index of bullishness of online posts can be defined on a daily basis according to \cite{Antweiler}:
%We need to aggregate texts features in order to compare hundreds of daily posts to the stock market volatility on a daily basis. If daily posts are ranked with binary labels, the bullishness is defined by
\begin{equation}
{B_t} = \ln \frac{{1 + N_t^p}}{{1 + N_t^n}}
\end{equation}
where ${N_t^p}$(${N_t^n}$) represents the number of positive (negative) posts on the day $t$. This indicator reflects both the expectations of the rise in price and the total number of posts.
When the posts have a continuous sentimental score instead of a binary label, the index of bullishness becomes
\begin{equation}
{B_t} = \ln \frac{{\varepsilon  + S_t^p}}{{\varepsilon  + \left| {S_t^n} \right|}} = \ln \frac{{\varepsilon  + \sum\limits_{k = 1}^{N_t^p} {{h_{\bf{w}}}{{({\bf{x}})}^{(k)}}} }}{{\varepsilon  + \left| {\sum\limits_{i = 1}^{N_t^n} {{h_{\bf{w}}}{{({\bf{x}})}^{(i)}}} } \right|}}
\end{equation}
where ${S_t^p}$(${S_t^n}$) represents the sum of positive (negative) sentimental score of the posts on the day $t$ and ${{h_{\bf{w}}}{{({\bf{x}})}^{(k)}}}$ (${h_{\bf{w}}}{{({\bf{x}})}^{(i)}}$) represents the positive (negative) sentimental score of the $k^{th}$ ($i^{th}$) post on the day $t$. $\varepsilon (\varepsilon>0)$ is a tiny number for smoothing, and we set $\varepsilon = 0.0001$ in our research.
The reason we use absolute value is because the score of negative sentiment is always less than zero.

\begin{figure}[htb]
\begin{centering}
\subfloat[]{\begin{centering}
\includegraphics[width=1\textwidth]{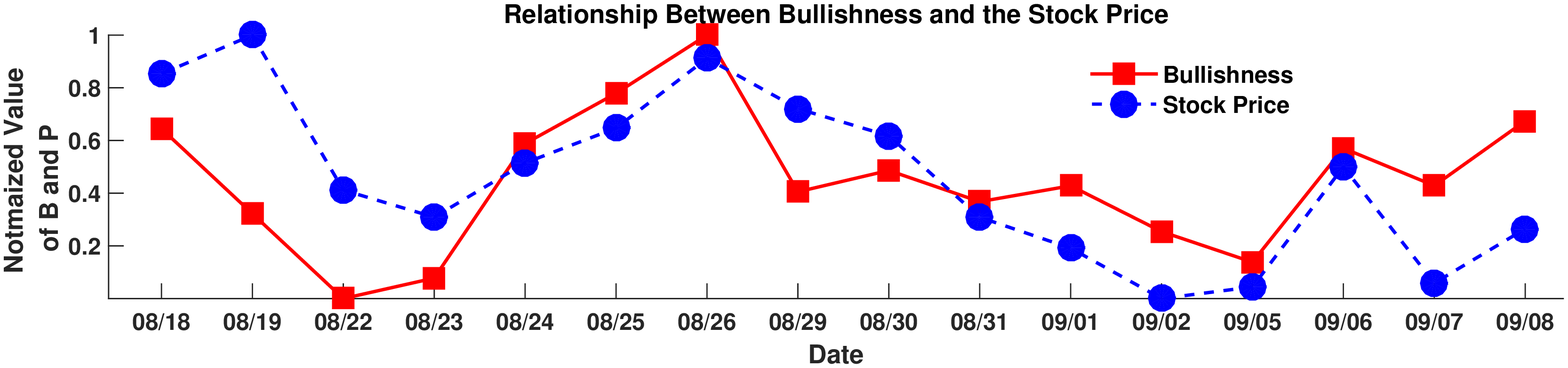}
\par\end{centering}
}

\subfloat[]{\begin{centering}
\includegraphics[width=1\textwidth]{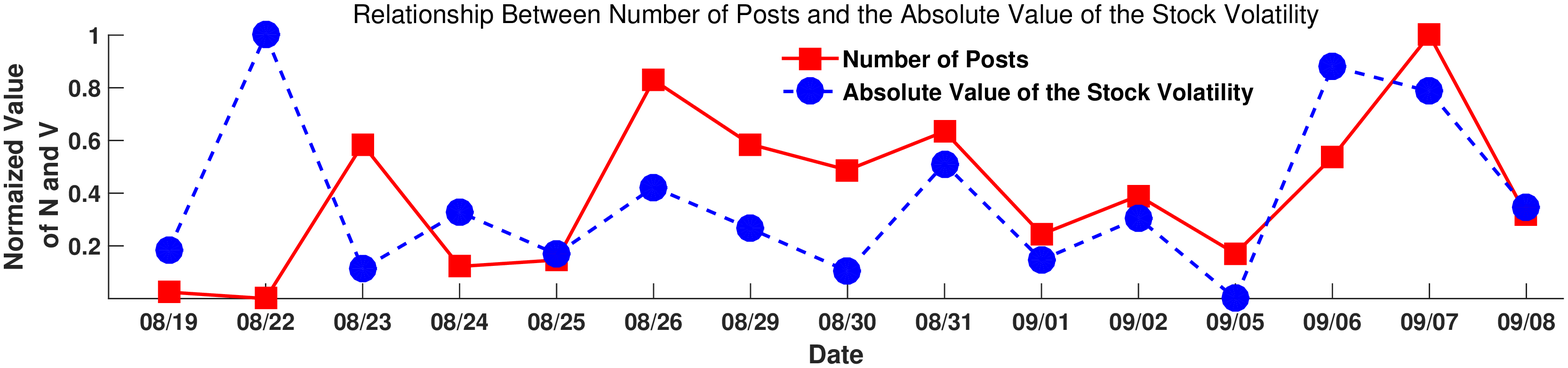}
\par\end{centering}
}
\par\end{centering}

\caption{\label{fig:r}Relationship between sentimental indicators and the stock information}
\end{figure}

The total number of the posts ${N_t}$ on the day $t$ is ${N_t}={N_t^p}+{N_t^n}$.
%it represent the posts volume as another sentimental indicators.
To enable fair comparison for ${B_t}$ and ${N_t}$, we use the z-score to normalize data based on the mean and standard deviation within a sliding window of length $l$ (we average the data of $l$ days before and after the current date $t$). The z-scores for ${{B_t}}$ and ${{N_t}}$ are:
\begin{equation}
{Z_{B}}^{(t)} = \frac{{{B_t} - \mu ({B_{t \pm l}})}}{{\sigma ({B_{t \pm l}})}}
\end{equation}
\begin{equation}
{Z_{N}}^{(t)} = \frac{{{N_t} - \mu ({N_{t \pm l}})}}{{\sigma ({N_{t \pm l}})}}
\end{equation}
%u是对于一个分布总体来说的，而x 拔是对于这一段的样本来说的，所以感觉这里还是用x 拔比较合适~
where ${\mu ({B_{t \pm l}})}$ and ${\mu ({N_{t \pm l}})}$ are the means and ${\sigma ({B_{t \pm l}})}$ and ${\sigma ({N_{t \pm l}})}$ are the standard deviations with $2l$ days around the current day $t$.
The correlations of $Z_B$ and stock price (Fig. 1-(a)), $Z_N$ and stock volatility (Fig. 1-(b)) given a particular stock are shown in Fig. \ref{fig:r}
We can see they are positively correlated and satisfy the two assumptions on sentimental indicators we previously gave.
%We normalize all the indicators sampled from the stock 700333 between 0 and 1. The price is the closing price while the stock volatility is the absolute value of the daily gains and declines. We can see that a higher bullishness brings out a rising price while a heated discussion in the forum will lead to a drastic fluctuation in prices. In order to study whether the relationship can help to guide the stock predict, we will make a comparison study in the next section.

\subsection{Stock Prediction with Recurrent Neural Network}
The stock price for a day is the weighted average price of all transactions on that day. In the Chinese market, it is
calculated by the last minute of the trading day, so it is also referred to as the closing price \cite{M2014Market}. In the actual stock market, profit-driving investors only care about the volatility of a stock instead of the exact price.
The stock volatility ${V_t}$ is defined based on the closing price $P_t$ on the current day $t$ and the previous day $t-1$:
\begin{equation}
{V_t} = \frac{{{P_t} - {P_{t - 1}}}}{{{P_{t - 1}}}};{V_t} \in [ - 0.1,0.1]
\end{equation}
%where ${{P_t}}$ refers to the closing price at day $t$.
%However, Chinese stock has a market characteristic to maintain the stability of the stock market, which stipulates the volatility should not exceed ten percent, otherwise the market will be shut down.
In order to regulate the stock market from any malicious manipulations, any stock has the volatility more than $10\%$ will be forced to quit the market on that trading day, therefore, ${{V_t}}$ always lies in the range of $[-0.1, 0.1]$.
%But this policy also curbs the impact of blind obedience on the stock price which will reduce our forecast accuracy.
In our experiment, we normalize the volatility into a time series between 0 and 1 based on mini-max normalization method, and generate the normalized time series $\bf{V}$. At the same time, we set 0.5 as the threshold in order to obtain a binary label (0 for price going down, and 1 for price rising).

\begin{equation}
{F_t} = \left\{ {\begin{array}{*{20}{c}}
1&{{{V}_t} > 0.5}\\
0&{otherwise}
\end{array}} \right.
\end{equation}
\begin{figure}[htb]
\centering
\includegraphics[width=.9\textwidth]{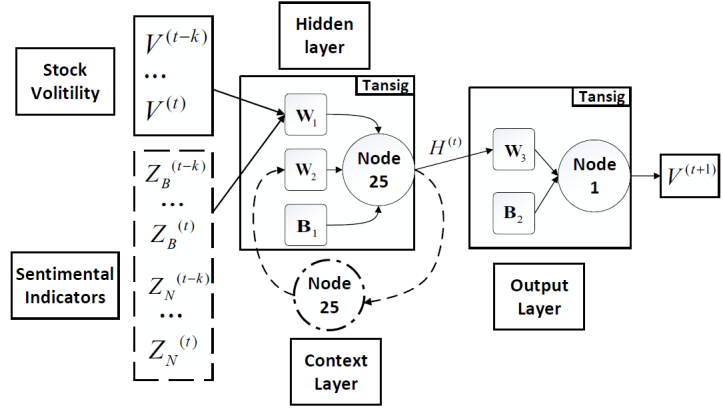}
\caption{The structure of the RNN model with sentimental indicators.
The input values are stock volatility ($V$) and sentimental indicators ($Z$),
we use the inputs of previous $k$ trading days to predict the stock volatility of the
next trading day ($V^{(t+1)}$). There are 25 hidden nodes used in our model.}
\label{fig:network}
\end{figure}

A stock market is highly complex in the control of ``invisible hands''.
However, there are still loads of research on statistical modeling and machine learning approaches to learn from history data.
The key to predict the stock market is to fit a latent nonlinear relation between the history data and the future stock volatility. The traditional statistical models used for financial forecasting were simple and suffered from several shortcomings. Machine learning methods like Multi-Layer Perception (MLP), Recurrent Neural Networks, Support Vector Machine (SVM) \cite{Rout2015Forecasting} have an increasing popularity in this area.% Unlike other feed forward neural networks, RNNs can use their internal memory to process arbitrary sequences of inputs.
%Considering the historical pattern, a window of inputs is prepared for the prediction model.
 %\cite{Mccluskey1994Feedforward}.
RNN is incorporated in our fundamental prediction model due to its appropriateness to address time series problem. The context layer stores the outputs of the state neurons from the previous time step and outputs to the next time step for computation.
In this paper, we employ Elman Network \cite{Elman1990Finding} in the following experiments.
%It uses context units to store the output of the state neurons from computation of the previous time steps. The context layer is used for computation of present states as they contain information about the previous states.
If we denote the output of hidden layer at time $t$ by $H^{(t)}$, the final prediction can be made by:
\begin{equation}
V^{(t+1)} = f(H^{(t)} W_3 +B_2)
\end{equation}
\begin{equation}
H^{(t)} = f( [V, Z]W_1 + H^{(t-1)} W_2 +B_1)
\end{equation}
where $f(\cdot)$ is the activation function and $B$ is bias. The structure of our proposed model \footnote{github link: \url{https://github.com/irfanICMLL/EMM-for-stock-prediction}.}  is show in Fig. \ref{fig:network}.

\section{Experimental Studies}
In order to verify the effectiveness of the new proposed model, we test it on the stock data introduced in Section 2.1.
The stock data of 250 consecutive trading days is downloaded from the DaZhiHui (DZH)\footnote{It can be downloaded from \url{http://www.gw.com.cn}.} software.
%The time
%We use a lagged time window with the length of $k$ and extract the historical data from the day $(t-k)$ to the day $t$, which
%we use for predicting the stock movements on the day $(t+1)$. We use the RNN introduced in the last section.
%there are $m$ indicators we want put into the predict model.
%The total input of the network will be ${m}\times{k}$.
%For a certain $k$ with $n$ days of historical data,
%we obtain $(n-k)$ pieces of data with window inputs and a single output.
%Each piece of data is independent from each other. We select the first 90\% as the training data, the latter 10\% as the test data. %
In order to evaluate the quality of the model, we define the concept of accuracy based on the binary label $F$. $F^*$ is the predict label of the test data while $F$ is the real label. Define $counter$ as the total number if ${F^*}_t = {F_t}$, accuracy 
%\begin{equation}
$Acc = \frac{{{counter}}}{{\left\| F \right\|}}$. 
%\end{equation}

\begin{figure}
\centering
\includegraphics[width=1\textwidth]{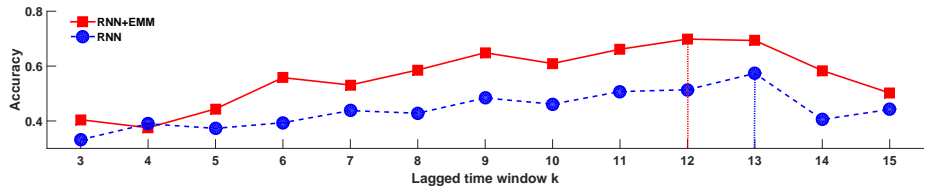}
\caption{Comparison results with different $k$. }
\label{fig:timeacc}
\end{figure}

We choose one stock (000573) as an example, we extract the volatility data and run RNN on it, and then compare to the RNN with sentimental indicators ${{Z}_{B}}^{(t)}$ and ${{Z}_{N}}^{(t)}$ (RNN+EMM).
%time series and converse them into RNN. The only difference between two control groups is whether or not they are added the sentimental time series
We vary $k$ from 3 to 15 to test the best length of history for predicting the future. The experiments are replicated for 50 times. Comparison results are shown in Fig. \ref{fig:timeacc}.
We can see that sentimental indicators help to improve the accuracy significantly, and the parameter $k$ will affect the prediction accuracy, the optimal length is around 10 based on different data sets. For the stock 000573, the best accuracy of the EMM with RNN is $69.85\% (k=12)$ while the best accuracy without sentimental information is $57.33\% (k=13)$, and the accuracy is significantly better than 0.5. Another 9 stocks are selected randomly to test the model, that increase the credibility of the conclusion.

\begin{figure}[htb]
\centering
\includegraphics[width=0.9\textwidth]{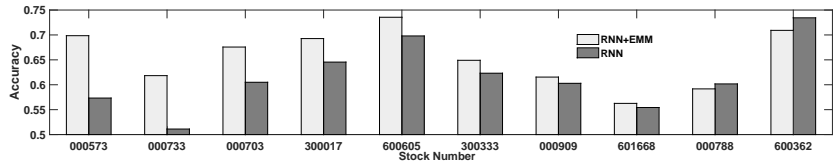}
\caption{Accuracy for RNN+EMM and RNN on 10-stocks dataset}
\label{total}
\end{figure}

As for each particular stock, we can obtain better performance for 8 datasets in 10 and the detailed result comparisons are shown in  Table. \ref{tab:result} . To make it more intuitive, we draw the histogram in Fig. \ref{total}. From the results, we can see that the stock 000573 performs better than others. The reason may be that most of the training posts of the emotion classifier come from its sub-forum during the chosen period. In other words, if the actual sentimental indicators are obtained, the accuracy of the model can be better.

\begin{table}[htb]
\centering
\caption{\label{tab:result} Accuracy and the best $k$ for RNN+EMM and RNN }
\begin{tabular}{ccccccccccc}
\hline
stock number& 000573 & 000733 & 000703 & 300017 &600605& 300333 &000909&601668&000788&600362 \\
\hline
RNN+EMM&0.6985&0.6187&0.6757&0.6927&0.7355&0.6491&0.6154&0.5626&0.5917&0.7092\\
RNN&0.5733&0.524&0.605&0.6455&0.6982&0.6232&0.6029&0.5543&0.6017&0.7344\\
$k$ (RNN+EMM) & 12&13&11&14&3&4&11&15&13&11\\
$k$  (RNN)    &13&5&5&14&14&6&10&3&13&6\\
\hline
\end{tabular}
\end{table}

Table \ref{tab:com} shows the comparison results of four learning models: MLP \cite{Frank2000Input}, SVM \cite{Burges1998A}, RNN, EMM+RNN and the baseline is a random guesser (RAND). On the 10 datasets with online discussions, the accuracy of RNN is higher than MLP and SVM,  because it contains information about the previous states. When considering sentimental indicators, the prediction performance  improves nearly 4\% on the average which verifies the assumptions about the sentimental indicators.
We only test 10 stocks, because the posts are not so easy to be obtained and labeled. We need roughly 20 hours to collect one sub-forum. And then extract the sentimental indicators through the method introduced in Section 2. In order to make the results more credible, we do repetition under various of the initial parameters to make sure that the improved accuracy is thus most likely not the result by chance nor the selection of a specifically favorable test period.

\begin{table}[!htb]
\centering
\caption{\label{tab:com}
Performance comparisons on 10 stocks from the Chinese market. }
\begin{tabular}{cccccc}
\hline
Method& RAND & MLP & SVM & RNN & RNN+EMM\\
\hline
MEAN&0.500168&0.559257&0.602339&0.61623&0.6549\\
STD& 0.003846& 0.028681& 0.111318& 0.06347&0.05640\\
\hline
\end{tabular}
\label{tab:com}
\end{table}

\section{Conclusions}
In this research, we investigated the relationship between the stock volatility and sentimental information obtained from an online stock forum. We employed a RNN model to consider sentimental
information, experimental results show that the new model can boost the prediction accuracy.
% and come to a conclusion that the information from the stock forums can help to predict the volatility of the stock market %to a certain degree. Furthermore, we dig out that the stocks that perform better on the new approach often have a lower accuracy %before we add the emotional information compared with those “badly performed ones” just like 600362. Although the accuracy %of stock 600362 decreases after using the new approach, it still has quite a high accuracy of 73.44\% with the old approach.
%In view of this phenomenon, we propose a reasonable hypothesis: When the stock price fluctuations conform to some law of history, change gently, or there are not any policy event occurs, the model performs well in prediction just according to the historical data. In the meanwhile, when the stock price skyrocket or plummet, or there were some social hot events associated with the stock spark a heated discussion among investors, it creates a new and unpredictable situation for historical data and lead to a low accuracy. However, the emotional indicators to a certain extent, contain this trend, and even lead to this trend. As a consequence, the emotional indicators improve the accuracy of the forecast effectively.\\
The main contribution of our research are as follows:
    (1) Generate a sentimental weight dictionary of Chinese stock posts.
    (2) Propose sentimental indicators and investigate the relationship between the stock volatility and the information from the stock forums.
    (3) Build a RNN model considering sentimental information for stock prediction and verifies the information from forums can help to predict the stock market of China. (4) We construct a benchmark dataset of labeled financial posts and make it public available for comparison studies.
Finally, it's worth mentioning that our analysis doesn't take into account many factors. The posts from the forums may contains a lot of fake messages that confuse the public. We will consider that in our future work. 

%And the accuracy of the simple linear emotion classifier is still improvable, because the corpus is very complex. After obtaining a larger data source, we can consider other word embedding model like word2vec. For the stock prediction, we have chosen a simple RNN model to study the effect of the sentimental index. This model can only predict that the stock will rise or not, but can not tell the range of increase or fall. We may consider more complex models like LSTM or GRU in our future research.

\section*{Acknowledgement}
This work is supported by the National Science Foundation of China Nos. 61401012 and 61305047.

\bibliographystyle{splncs03}
\bibliography{llncs}

\end{document}